\documentstyle[floats,multicol,aps,epsf,epsfig,prb]{revtex}  

\newcommand\dg{^{\dag}}  
\newcommand\la{{\langle}}  
\newcommand\ra{{\rangle}}  
\begin{document}  
\ifx\href\undefined\else\errmessage{hyperTeX disabled by xxx admin}\fi  
\draft  
\twocolumn[\hsize\textwidth\columnwidth\hsize\csname @twocolumnfalse\endcsname  
\title{Gap-anisotropic model for the narrow-gap Kondo insulators}  
\author{Juana Moreno$^{1}$ and P. Coleman$^{2}$}  
\address{$^{1}$ Dept of Physics \& Astronomy, Northwestern Univ,2145 Sheridan Rd, Evanston IL 60208, }  
\address{$^{2}$Center for Materials Theory,
Department of Physics and Astronomy, 
Rutgers University, Piscataway, NJ 08854, USA.
}
\date{\today}  
\maketitle  
\begin{abstract} 
A theory is presented which accounts for the dynamical  
generation of a hybridization gap with nodes in the Kondo insulating  
materials $CeNiSn$ and $CeRhSb$.  
We show that Hunds interactions acting on virtual   
$4f^2$ configurations of the cerium ion  
can act to dynamically select the shape of the cerium ion 
by generating  
a Weiss field which couples to the shape of the ion.   
In low symmetry crystals where the external crystal fields  
are negligible, this process  
selects a nodal Kondo semimetal state as the lowest 
energy configuration.  
\end{abstract}  
\pacs{Pacs numbers:}  
\vskip2pc]  
  
\def\beq{\begin{equation}}  
\def\eeq{\end{equation}}  
\def\bea{\begin{eqnarray}}  
\def\eea{\end{eqnarray}}  
\def \vk{\vec k}  
\def \rarrow{ \rightarrow}  
  
Kondo insulators share in common with the   
Mott insulators a gap which is driven by  
interaction effects.\cite{fisk,ueda}  
Unlike Mott insulators, they  undergo a   
smooth cross-over into the insulating state,  
where a tiny charge and spin gap develops.  
These materials  
are generally regarded as a  special   
class of heavy fermion system, where  
a lattice  Kondo effect between the  
localized spins and  conduction electrons   
forms a highly renormalized band-insulator.  
\cite{bedell,riseborough}  
  
The smallest gap Kondo insulators, $CeNiSn$ and $CeRhSb$,  
do not naturally fit into this   
scheme: they appear to develop  
gapless excitations.  
Early measurements showed a drastic increase of the electrical resistivity  
below $6 K$,\cite{CeNiSn-thermal1} but very pure samples of   
$CeNiSn$   
display metallic behavior.\cite{CeNiSn-conduc}  
NMR measurements   
are consistent with an electronic state  
with a ``v-shaped'' component to the density of states.\cite{1Kondo-NMR}  
These results, together with   
other transport properties   
\cite{CeRhSb-thermal,CeNiSn-thcond,CeNiSn-neutron-old,CeNiSn-neutron-new}  
point to the formation of a new kind of semi-metal with an  
anisotropic hybridization gap.  
  
Ikeda and Miyake \cite{Miyake} (IM) recently   
proposed that  the Kondo insulating ground-state of these  
materials develops  
in a  crystal field state with an axially symmetric  
hybridization potential that vanishes along a single  
crystal axis.  This picture  
accounts for the v-shaped density of states, and provides an  
appealing way to understand   
the  anisotropic transport at  
low temperatures, but it leaves a number of    
puzzling questions.   
In $CeNiSn$ and $CeRhSb$, the  Cerium ions are located  
at sites of minimal monoclinic symmetry, where the low-lying  
f-state is a Kramers doublet   
\bea  
\vert \pm \rangle =   
b_1\vert \pm 1/2\rangle + b_2 \vert \pm 5/2\rangle  
+b_3 \vert \mp 3/2\rangle  
\eea  
where $\hat b= (b_1,b_2,b_3)$ could point anywhere on  
the unit sphere, depending on details of the monoclinic crystal field.  
The IM model  
corresponds to three symmetry-related 
points in 
the space of crystal field ground states,  
\bea 
\hat b= \left\{ \begin{array}{lr} 
(\mp \frac{\sqrt{2}}{4},-\frac{\sqrt{5}}{4},\frac{3}{4}) \cr 
(0,0,1) 
\end{array} 
\right. 
\eea 
where a node develops 
along the $x$, $y$ or $z$ axis respectively.  
What mechanism selects  
this special semi-metal out of the manifold  
of gapped Kondo insulators?   
Neutron  scattering results show  no  
crystal  
field satellites in the dynamical spin susceptibility  
of CeNiSn, \cite{Alekseev}  suggesting the   
the crystal electric fields are quenched: 
is the selection of the nodal semi-metal   
then a many body effect?\cite{Prokof'ev}   
  
In this letter, we propose that   
this selection mechanism  
is driven by  
Hund's interactions amongst f-electrons  
in the Cerium ions.   
Hund's interactions play an important role in multi f-electron  
ions. \cite{norman}  
In the Kondo semi-metal, the Cerium ions are in  
a nominal $4f^1$ state, but   
undergo   
valence fluctuations  into $f^0$ and $f^2$ configurations.  
We show that the memory effect of the Hund's interactions 
in the $f^2$ state induces a kind of Weiss field which couples to the   
shape of the Cerium ion.  When this field adjusts to minimize  
the Hund's interaction energy, the nodal IM state is selected.

To develop our model, we classify each  
single-particle f-configuration by a ``shape'' (a=1,2,3) and a pseudo-spin  
quantum number($\alpha  = \pm 1$), where  
\bea  
f\dg_{1\pm}\vert 0\rangle &\equiv& \vert \pm 1/2\rangle,\cr  
f\dg_{2\pm}\vert 0\rangle &\equiv& \vert \pm 5/2\rangle, \cr  
f\dg_{3\pm}\vert 0\rangle &\equiv& \vert \mp 3/2\rangle  
\eea  
There are eight  
multipole operators   
\bea  
[\pmb{$\Gamma$}]^a=  f\dg_{b  
\sigma}\Lambda^a_{b, c}  f_{c \sigma}, \qquad (  
a=1,8)  
\eea  
which describe the shape of the Cerium ion,   
where   
the $\Lambda^a$ matrices are the eight traceless SU(3) generators,
normalized so that ${\rm Tr}[\Lambda^a \Lambda^b]= \delta^{ab}$, 
We shall describe the low energy physics   
by an Anderson  model  $H=H_o+H_f$, where 
\bea  
H_o=H_c +  
\sum_{ ja \sigma}  
V[ c\dg_{a \sigma}(j)f_{a \sigma}(j)   
+ {\rm H.c} ],\eea  
and $H_c=\sum_{{\bf k} \sigma}\epsilon_{\bf k} c\dg_{{\bf k} \sigma}  
 c_{{\bf k} \sigma}  
$  
describe a spin-1/2 conduction band hybridized with a lattice of localized  
f-states. The operator   
\bea  
c\dg_{a \alpha}(j) =  (N_s)^{-\frac{1}{2}}\sum_{\bf k,  
\sigma} e^{-i {\bf k}\cdot {\bf R_j}}  
{\cal Y}_{a \alpha}^{\sigma}({\bf \hat k}) c\dg_{\bf k \sigma}  
\eea  
creates a conduction electron   
in a $l=3$, $j=5/2$   
Wannier state at site $j$ with shape-spin quantum numbers $(a,\   \sigma)$, 
$N_s$ is the number of sites and   
\bea  
{\cal Y}_{a \alpha}^{\sigma}({\bf \hat k})=   
Y^{m_J- \sigma}_3 ({\bf \hat k}) ({\textstyle \frac{1}{2}}\   
\sigma,3\ m_J-\sigma  
\vert {\textstyle \frac{5}{2}} m_J)  
\eea  
defines the form-factors,   
in terms of spherical  
harmonics and the Clebsh-Gordon coefficients   
of the $j=5/2$ $f^1$-state\cite{angular},    where 
$m_J\equiv m_J(a,\alpha)$  
maps the spin-shape quantum numbers to to the original  
azimuthal quantum number of the f-scattering channel.  
Following previous authors, \cite{read}  
we regard $H$ as a low energy Hamiltonian, so  
that hybridization strength  
$V$ is a renormalized quantity, that takes into account  
the high energy valence and spin fluctuations. 
 
The term   
\bea  
H_f=\sum_j  
\bigl[ E_f n_f(j)+  
\frac{U}{2}(n_f(j)-1)^2 -\frac{g}{2}\pmb{$\Gamma$}_j^2\bigr]  
\eea  
describes the residual low-energy interactions amongst the 
f-electrons:  
the second term is a Coulomb  
interaction term. The   
third term is  
a Hunds interaction which  
favors $4f^2$ states with maximal total  
angular momentum. In  
an isotropic environment, this interaction 
would take the form $- \frac{g}{2} { \bf J}^2$, where ${\bf J}$ is the total 
angular momentum operator, but in a crystalline  
environment, it takes on  a reduced symmetry 
which we model in simplified form by  
$- \frac{g}{2} {\pmb{$\Gamma$}}^2$.  
In general the Hund's interaction  
is only invariant under discrete rotations so that fluctuations  
into the $f^2$ state enable the system to sample the crystal  
symmetry even when the conventional crystal field splittings 
are absent.  
 
Suppose the  
crystal electric field term were unquenched, so that 
$H\rarrow H  
- \sum   
\pmb{$\alpha$}\cdot \pmb {$\Gamma$}_j$.  
The shape of  the Cerium ion   
 $  
\la \pmb{$\Gamma$}_j\  
\ra= \pmb{$\Gamma$}$ is determined by the condition  
that the energy is 
stationary with respect to  
variations in  
$\pmb{$\Gamma$}$,   
\bea  
N_s^{-1}{\delta \la H_o\ra  
}/{\delta \pmb{$\Gamma$}}  
=  \pmb{$\alpha$}+g\pmb{$\Gamma$}.  
\eea  
The second term is a  feedback or ``Weiss'' contribution   
to the crystalline electric field, created by fluctuations  
into the $4f^2$ state. Generally, 
the induced field $\Gamma$ will follow the crystalline 
electric fields $\pmb{$\alpha$}$, but in situations  
where the valence and spin fluctuations are  
rapid enough to quench the external crystal electric field,\cite{Alekseev}   
then $\pmb{$\alpha$}=0$,  
and the Weiss field  
becomes  
free to explore phase space  
to minimize the total energy.   
In such a situation,  
the shape of the Cerium ion is   
determined by the interactions,  
rather than the local conditions around each ion.   
  
To explore  
this process, we carry out a Hubbard-Stratonovich decoupling of the   
interactions,  
\bea  
H_f(j)\rarrow  f\dg_j  
\bigl[(E_f+\lambda_j)\underline{1} +\pmb{$\Delta$}_j 
\cdot \underline{\pmb{$\Lambda$}}\bigr]  
f_j + E_o[ \lambda_j, \pmb{$\Delta$}_j]\label{xtalf}  
\eea 
where 
\bea 
E_o[ \lambda_j, \pmb{$\Delta$}_j]=  
\bigl(\frac{\pmb{$\Delta$}_j  ^2}{2g} -\frac{\lambda_j^2}{2U}-\lambda_j 
\bigr),  
\eea   
Here  
$\pmb{$\Delta$}^a(j)\sim -g\pmb{$\Gamma$}^a(j)$ is a dynamical Weiss  
field,  
$f_j$ denotes the spinor  
$f_j\equiv f_{a \sigma}(j)$.  
Note that in the  path integral,   
the  
fluctuating part of $\lambda_j$,  
associated with the suppression of charge fluctuations, is imaginary.  
We now seek a mean-field solution where  the Weiss field  
$\lambda_j=\lambda$ and $\pmb{$\Delta$}_j = \pmb{$\Delta$}$, and 
$E(\lambda_j, \pmb{$\Delta$}_j) = E_o$.  
Such an expectation value does not break the crystal symmetry.  
However,  
the selected crystal field matrix $\pmb{$\Delta$}\cdot \pmb{$\Lambda$}$  
must adjust to minimize the total energy.   
Supposing we diagonalize this matrix,   
writing  
 $  
\pmb{$\Delta$}  
\cdot \underline{\pmb{$\Lambda$}}= U \underline{\pmb{$\Delta$}_o}U\dg$, where  
$  
\underline{\pmb{$\Delta  
$}_o}= {\rm diag}(  
\Delta_1,  
\Delta_2,  
\Delta_3)  
$, and  
$\Delta_1>\Delta_2> \Delta_3$.  
In the  basis, $\tilde f_{a\sigma}(j) = U\dg_{ab}   
f_{b\sigma}(j)$, the crystal field is diagonal.  
In practice, the strength of the Hund's interaction  
$g$ is so large  that   
the excitation energies $\Delta_{1,2}-\Delta_3$   
substantially exceed the Kondo temperature. In this case,  
the  
mean-field Hamiltonian must be projected into the subspace of the lowest  
eigenvalue. In the hybridization, we therefore  
replace   
\bea  
c\dg_jf_j= c\dg_j \underline {U}  
\tilde f_j\rarrow b_{a} [ c\dg_{a\sigma}(j)\tilde f_{\sigma}(j)],  
\eea   
where   
$\tilde f_{\sigma}(j)\equiv \tilde f_{3 \sigma}$  
(dropping the superfluous index ``3'')   
describes the lowest Kramers doublet and   
$b_a\equiv U_{a3}$.   
To satisfy the constraint $\la n_f\ra =1$, the energy of  
the lowest Kramers doublet must be zero, i.e.   
$E_f+\lambda + \Delta_3=0$. We then arrive at the  
mean-field Hamiltonian  
\bea  
H^*=H_c  
+ V\sum_{\bf k} [\phi_{\sigma \alpha}({\bf k}) c\dg_{{\bf k}\sigma}f_{\alpha \bf k}  
+ {\rm H.c} ]+N_sE_o  
\label{effect-Anderson}  
\eea  
where $\phi_{\sigma \alpha }({\bf k}) = \sum_a b_a {\cal Y}_{a \alpha}^{\sigma}  
({\bf \hat k})$  
is the dynamically generated form-factor of the hybridization.\cite{angular} 
The transformed 
hybridization is no longer rotationally invariant: 
all information about the anisotropic wavefunction 
of the Cerium ion is now encoded  
in the vector $\hat {\bf b}$.  
 
The quasiparticle energies associated with this Hamiltonian are  
\bea  
E^{\pm}_{\bf k} = \epsilon_{\bf k}/2 \pm \sqrt{  
(\epsilon_{\bf k}/2)^2 + V_{\bf k}^2 }  
\label{hyb-kondo}  
\eea  
Here, the hybridization can be written in the convenient form  
$  
V_{\bf k}^2= V^2 \Phi_{\hat b}({\bf  \hat k})  
$  
where $\Phi_{\hat b}({\bf \hat k})  
=(1/2) \sum_{\alpha, \sigma} \vert \sum_a b_a {\cal Y}_{a \alpha}^{\sigma}  
({\bf \hat k}) \vert ^2 $ contains all the details of the gap  
anisotropy.  
The ground-state energy is then  
the sum of the energies of the filled lower band  
\bea  
E_g = -2 \sum_{\bf k} \sqrt{  
(\epsilon_{\bf k}/2)^2 + V_{\bf \hat k}^2} + N_sE_o  
\label{precise}\eea  
Now both $\lambda$ and $\Delta_3$ are fixed independently of the  
direction of $\bf \hat b$, so that $E_o$ does not depend  on $\bf \hat b$.   
To see this, write the eigenvalues of the traceless 
crystal field matrix as  
$\Delta_{1,2} = \frac{1}{\sqrt{6}} \Delta \pm \delta$, $\Delta_3=  
- \frac{2}{\sqrt{6}}\Delta$.  
Since the upper two crystal field states are empty, stationarity w.r.t 
to $\delta$ requires $\delta=0$. 
Since $\Delta_3$ couples directly to the  
f-charge, we obtain $\partial E_g/\partial \Delta = -\sqrt{\frac{2}{3}}\la n_f\ra +  
(\Delta/g) 
=0$,  
so that $\Delta= \sqrt{\frac{3}{2}}g $.  Thus both  
$\lambda= -\Delta_3- E_f$ and $\underline{\pmb{$ \Delta$}}_o$ are fixed 
independently of $\hat b$.  
The selection of the crystal field configuration  
is thus entirely determined by minimizing the kinetic  energy of the 
electrons.  
\begin{figure}[t]  
\centerline{\epsfig{figure=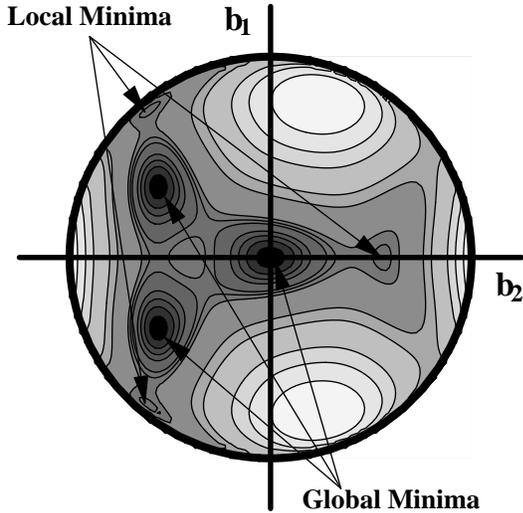,width=0.8\linewidth}}  
\caption[]  
{Contour plot of the ground-state energy in mean-field theory  
as a function of the two first components of the unit vector  
$\hat b$ (the third one is taken as positive).  
The darkest regions correspond to lowest values of the free-energy.  
Arrows point to the three global and three local minima.}  
\label{fig:free}   
\end{figure}  
To  examine the dependence of the mean-field on  
$\hat b$,  
we replace   
the momentum sum in  (\ref{precise})  
by an energy and angular integral,  
\bea  
\sum _{\bf k}\{\dots\}\rightarrow N(0)\int_{-D}^D  
d \epsilon  
\frac{d \Omega _{{\bf \hat k}}}{4 \pi}\{\dots\},  
\eea  
where $N(0)$ and $2D$ are respectively, the density of states and band-width 
of the conduction band.  
Completing the integral, noting that  the angular average  
$ \langle \Phi_{\bf \hat b}({\bf k}) \rangle = 1$, we find that  
the shift in the  ground-state energy per site due to the hybridization is 
\bea  
\Delta E_g  = 2 N(0) V^2 \biggl[  
{\rm ln } [\frac{V^2} {e D^2}] + F[ \hat b] \biggr]  
\eea  
where  
\bea  
F[\hat b] = \int \frac{d \Omega _{\bf \hat k}}{4 \pi}   
\Phi_{\hat b} ( {\bf \hat k})  
{\rm ln } \bigl[  
\Phi_{\hat b} ( {\bf \hat k})  
\bigr]  
\eea  
The weak logarithmic divergence inside $F({\bf \bf b})$ favors 
states with nodes.  
Fig. 1. shows a contour plot of the mean-field free-energy  
as a function of the two first components of $\hat b$.  
There are three global minima and three local minima with a slightly higher  
free-energy.  
The state where  $\hat b = \hat z$,   
plus two symmetry equivalents, corresponds  to the   
IM state and has the lowest free-energy.     
The IM state is axially symmetric, with a hybridization node along   
the $\hat z$, $\hat y$ or $\hat x$   axis.  
But the theory also identifies a new locally stable state  
where $\hat b =(0,\sqrt{5}/4,\sqrt{11}/4)$, plus its two symmetry  
equivalents. This state is   
almost octahedral.  Like the IM state,   
the hybridization drops exactly to zero   
along the $\hat z$ axis. But, in marked difference with the IM state,  
it almost vanishes along the $(1,1,0)$  
and $(1,-1,0)$ directions in the basal plane.    
  
\begin{figure}  
\noindent  
\center  
\begin{minipage}[tb]{0.8\linewidth}  
\centering{\epsfig{figure=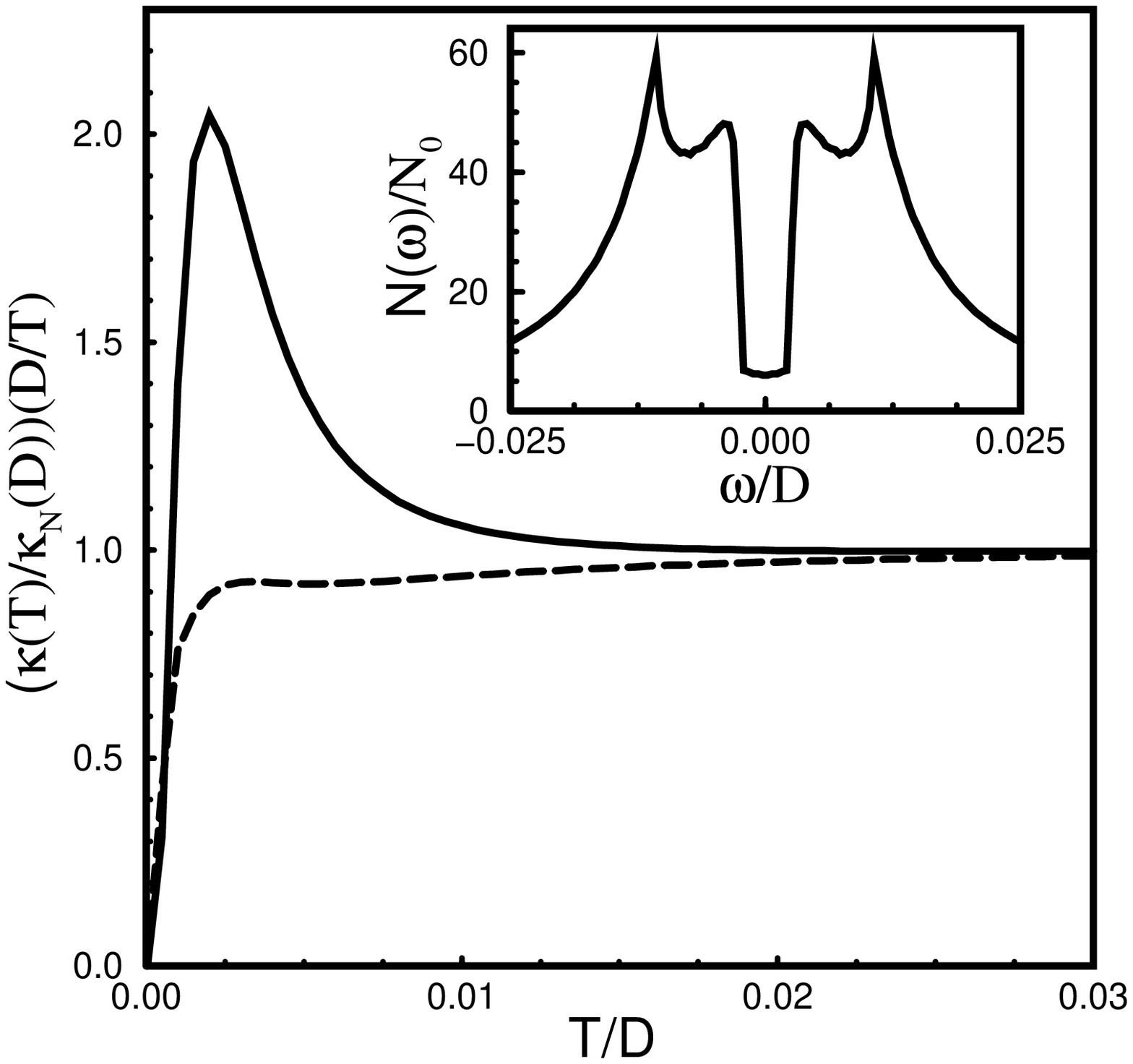,width=\linewidth}}  
\end{minipage}  
\begin{minipage}[t]{0.8\linewidth}  
\centering{\epsfig{figure=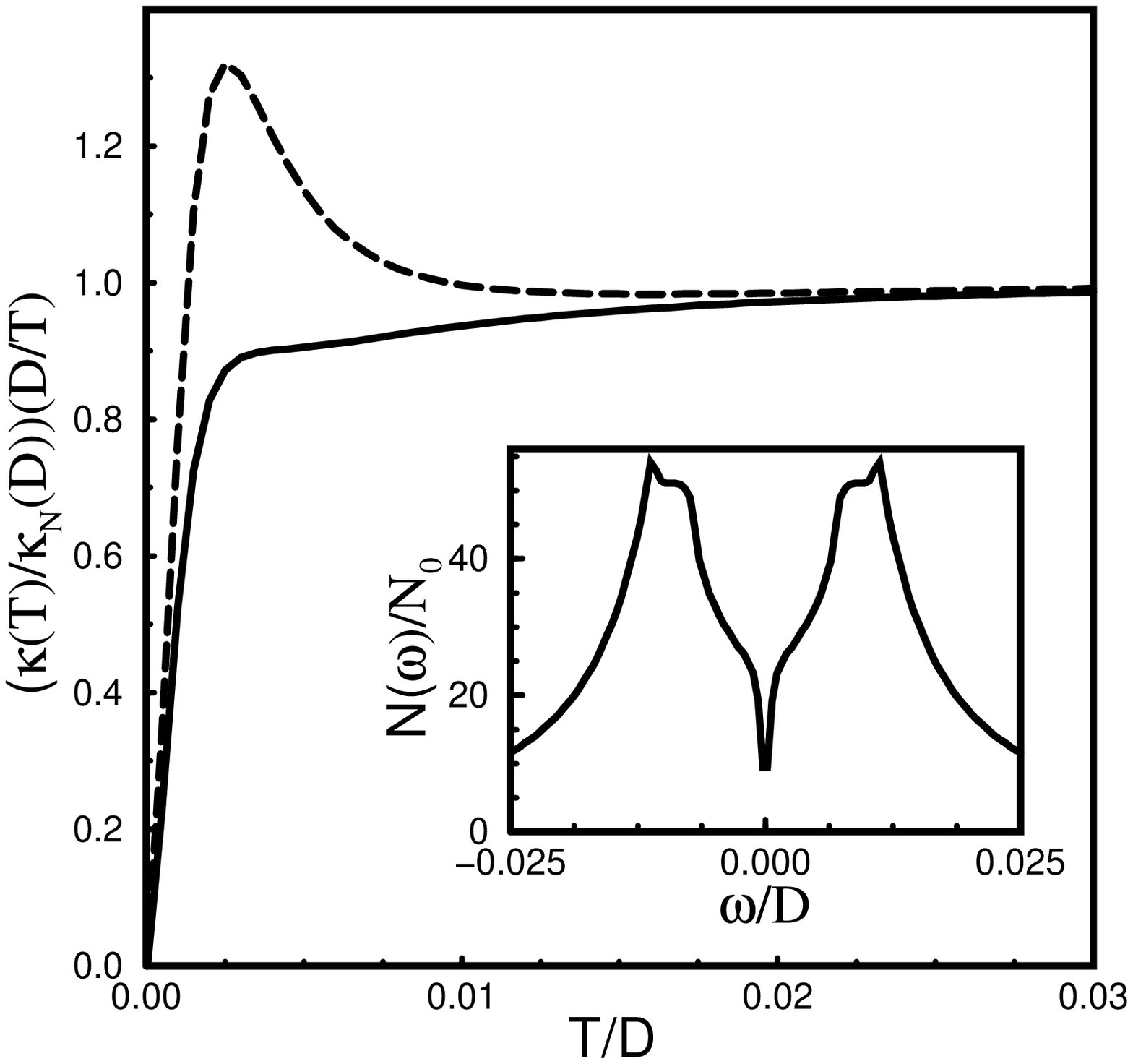,width=\linewidth}}  
\end{minipage}  
\caption[]   
{Normalized thermal conductivity   
versus temperature along the z-axis (solid line)    
and in the basal plane (dashed line). Top: for the Ikeda-Miyake state.   
Bottom: for the quasi-octahedral scenario.  
Insets show density of  
states as a function of the energy.  
The adjustable parameters have been chosen as  
$V/D=0.08$ and an impurity scattering  
phase-shift of $\pi/2$.}  
\label{fig:therm}   
\end{figure}

The relative stability  
of the IM and the octahedral state will, in general be dependent  
on details of our model, such as the detailed conduction electron  
band-structure. For this reason, both possibilities should be considered  
as candidates for the nodal semi-metallic states of   
$CeNiSn$ and $CeRhSb$.  
The inset in Fig (\ref{fig:therm}) shows the density of states  
predicted by these two possibilities. Although both are gapless,  
the v-shaped pseudogap of the quasi-octahedral state is far more pronounced  
than in the axial state, and is closer in character to the  
observed tunneling density of states. \cite{tunnel}  
A more direct probes of the anisotropy  
is provided by the   
thermal conductivity\cite{new.therm}  which, unlike   
the resistivity, does not show a strong sample dependence in  
these compounds.   
To compute and compare the theoretical thermal conductivity  
with experiments, we compute the thermal  
current correlator \cite{thermalcurrents}  
\begin{equation}  
\kappa^{ij} = { 1 \over 2T} \int _{-\infty}^{\infty}  
d \omega  
\omega^2\biggl(  
-{\partial f \over \partial\omega}  
\biggr){ N(\omega) \over \Gamma(\omega)}  
\langle \vec {\cal V}^i\vec {\cal V}^j \rangle_{\omega}  
,  
\end{equation}  
where $f$ is the  Fermi function,  
$\Gamma(\omega)$ is the quasiparticle scattering rate  
and  
\begin{equation}  
N(\omega)\langle \vec {\cal V}^i\vec {\cal V}^j\rangle_{\omega}  
 =\sum_{\vec k}  
\vec {\cal V}_{\vec k}^i  
\vec {\cal V}_{\vec k}^j \delta(\omega- \vec E_{\vec k})  
\end{equation}  
describes the quasiparticle velocity distribution where  
${\vec {\cal V}}_{\vec k}= \vec{\nabla}_{\vec k} E_{\vec k}$   
and $E_{\vec k}$ is   
given by equation (\ref{hyb-kondo}).   
For our calculation, we have considered   
quasiparticle scattering off   
a small, but finite density of unitarilly scattering impurities or   
``Kondo holes''.\cite{Schlottmann}  
We  use  a  self-consistent T-matrix approximation, following the lines  
of earlier calculations except for one key difference.  
In these calculations, which depend critically on the  
anisotropy,   
it is essential to include the momentum dependence of the   
hybridization potential in the evaluation of the   
quasiparticle current.  
Previous calculations \cite{Miyake} underestimated the anisotropy by  
neglecting these contributions.\cite{thermalcurrents}   
  
The single node in the IM state leads to a   
pronounced enhancement of the low-temperature  
thermal conductivity along the nodal $\hat z$   
axis. By contrast, in the quasi-octahedral state   
the distribution of minima in the gap give rise to a modest  
enhancement of the thermal conductivity   
in the basal plane.  
Experimental measurements \cite{new.therm} tend to favor the latter  
scenario, showing an   
enhancement in thermal conductivity that  
is much more pronounced in $\kappa_x$ than in   
$\kappa_z$ or $\kappa_y$.

Three aspects of our theory deserve more extensive  
examination. Nodal gap formation is apparently unique to  
$CeNiSn$ and $CeRhSb$; the   
other Kondo insulators $SmB_6$, $Ce_3Bi_4Pt_3$ and $YbB_12$ display a well-formed gap. Curiously, these materials are cubic, leading us   
to speculate that their higher symmetry prevents the dynamically generated  
contribution to the crystal field from exploring the region  
of parameter space where a node can develop.   
At present, we have not included the effect of a  
magnetic field, which is known to suppress the gap  
nodes.\cite{CeNiSn-conduc}    
There appears to be the interesting possibility  
that an applied field will actually modify the dynamically generated crystal  
field to  eliminate the nodes.   
Finally, we note that since the spin-fluctuation spectrum 
will reflect the nodal structure, future neutron scattering  
experiments\cite{CeNiSn-Aeppli} should in principle be  
able to resolve the axial or octahedral symmetry of the low energy  
excitations.  
  
To conclude,  we have proposed a mechanism for the dynamical  
generation of a hybridization gap with nodes in the Kondo insulating  
materials $CeNiSn$ and $CeRhSb$.    
We have found that Hunds interactions acting on the virtual  
$4f^2$ configurations of the Cerium ions generate a Weiss  
field which acts to co-operatively  
select a semi-metal with nodal anisotropy.   
Our theory predicts   
two stable states, one axial , the other quasi-octahedral in symmetry.  
The quasi octahedral solution appears to be the most  
promising candidate explanation of the various   
transport and thermal properties  
of the narrow-gap Kondo insulators.

We are grateful to Gabriel Aeppli, Frithjof Anders,  
Yoshio Kitaoka, Toshiro Takabatake and Adolfo Trumper  
for enlightened discussions.  
This research was partially supported by NSF grant DMR 96-14999 
and DMR 91-20000 through the 
Science and Technology Center for Superconductivity.
JM  acknowledges also support by the Abdus Salam ICTP.

\end{document}